\documentclass[twocolumn,10pt]{asme2e}

\usepackage{amsfonts}
\usepackage{amsmath}
\usepackage{xcolor}
\usepackage{listings}
\usepackage{float}
\usepackage{graphicx}

\definecolor{backgroundColour}{rgb}{1,1,1}

\lstdefinestyle{CStyle}{
    backgroundcolor=\color{white},   
    commentstyle=\color{black},
    keywordstyle=\color{black},
    numberstyle=\tiny\color{black},
    stringstyle=\color{black},
    basicstyle=\footnotesize,
    breakatwhitespace=false,         
    breaklines=true,                 
    captionpos=b,                    
    keepspaces=true,                 
    numbers=left,                    
    numbersep=5pt,                  
    showspaces=false,                
    showstringspaces=false,
    showtabs=false,                  
    tabsize=2,
    language=C
}
\confdate{May 3}
\confyear{2024}

\title{Parallelization of the K-Means Algorithm with Applications to Big Data Clustering}

\author{Ashish Srivastava
    \affiliation{
	ME21B036 \\
 Indian Institute of Technology Madras
    }	
}

\author{Mohammed Nawfal     
    \affiliation{ME21B046 \\
    Indian Institute of Technology Madras
    }
}

\author{Course Instructor:    
    \affiliation{Prof. Kameswararao Anupindi
    }
}

\begin{document}

\maketitle    

\begin{abstract}
{\it The K-Means clustering using LLoyd's algorithm is an iterative approach to partition the given dataset into K different clusters. The algorithm assigns each point to the cluster based on the following objective function
\[\  \min \Sigma_{i=1}^{n}||x_i-\mu_{x_i}||^2\]
The serial algorithm involves iterative steps where we compute the distance of each datapoint from the centroids and assign the datapoint to the nearest centroid. This approach is essentially known as the expectation-maximization step.
\par
Clustering involves extensive computations to calculate distances at each iteration, which increases as the number of data points increases. This provides scope for parallelism. However, we must ensure that in a parallel process, each thread has access to the updated centroid value and no racing condition exists on any centroid values.
\par
We will compare two different approaches in this project. The first approach is an OpenMP flat synchronous method where all processes are run in parallel, and we use synchronization to ensure safe updates of clusters. The second approach we adopt is a GPU based parallelization approach using OpenACC wherein we will try to make use of GPU architecture to parallelize chunks of the algorithm to observe decreased computation time.  We will analyze metrics such as speed up, efficiency,time taken with varying data points, and number of processes to compare the two approaches and understand the relative performance improvement we can get. .}
\end{abstract}

\begin{nomenclature}
\entry{$\mathbf{X}$}{The dataset on which clustering is to be performed.}
\entry{$\mathbf{x}_i$}{A d-dimensional vector.}
\entry{$N$}{The size of the dataset.}
\entry{$z_i^t$}{The cluster indicator of $\mathbf{x}_i$ at the $t^{th}$ iteration. Can take any integer value from 1 to K.}
\entry{$\mu_k^t$}{The cluster center of the $k^{th}$ cluster during the $t^{th}$ iteration.}
\entry{$\mathbf{1}(z)$}{The proposition function.}
\entry{$\mathbf{E}$}{The L2 norm error between the cluster centers of two successive iterations.}
\end{nomenclature}

\section*{INTRODUCTION}

Clustering is an unsupervised machine-learning technique used for grouping data in such a way that similar samples are clustered into the same group due to some underlying pattern without any explicit markers or labeling. Clustering finds applications in various fields such as pattern recognition, image segmentation, anomaly detection, and data compression.

Various clustering algorithms exist, with each having its metric in defining similarity, partitioning of data, and quality of clustering. Some of the popular clustering algorithms that are widely used include K-Means, Hierarchical clustering, DBSCAN, and Gaussian Mixture Models.

The K-Means clustering algorithm is one such clustering algorithm that makes use of an iterative method in clustering N data points into K clusters by computing the cluster mean for every cluster and minimising the L2 distance of each point from the cluster centers/means during each iteration. The time complexity for the algorithm is $O(NTKd)$, where N is the total number of data points, T is the number of iterations required for the algorithm to converge, K is the number of clusters, and d is the dimension of each data point. Since the values of K and d are generally way smaller than N, they can be ignored. It is also observed that $T \propto N$, hence the effective time complexity reduces to $O(N^2)$. Due to this high time complexity, the computation of the clusters would be time-consuming for large datasets.

However, due to the simple nature of the steps and calculations in the algorithm, it provides us with the scope of running the algorithm in a parallel or distributed environment. This paper aims to explore the scope and extent of parallelisation of the algorithm mainly using two parallelisation models, namely a Shared Memory Model using OpenMP and a GPU Programming Model using OpenACC.

\section*{PROBLEM STATEMENT}
Given an extensive dataset comprising data points in a two and three-dimensional space, the aim is to cluster these data points into K clusters using the Lloyd's Algorithm for K-means clustering. The algorithm is to be appropriately parallelised based on the parallelisation paradigm chosen and the relevant parallelisation parameters are to be evaluated to compare performances. 

\section*{THE LLOYD'S ALGORITHM}

The Lloyd's algorithm is an iterative method that forms the basis of K-Means clustering. Given a dataset $\mathbf{X} \in \mathbb{R}^{d \times N}$, where $\mathbf{x}_i$ is a d-dimensional vector, the Lloyd's algorithm can be broken down into three steps as follows.

\begin{enumerate}
    \item Initialisation: Assuming the dataset is to be clustered into K clusters, we first initialise the cluster centers by randomly selecting K points from the dataset. These initial cluster centers are denoted by $\mu_1^0,\mu_2^0,\mu_3^0, \cdots,\mu_K^0$, where the superscript indicates the iteration.
    \item Reassignment: For the $t^{th}$ iteration, the distance of every $\mathbf{x}_i$ from $\mu_k^t$, is computed for all values of k. If $z_i^t$ denotes the cluster indicator of $\mathbf{x}_i$ for the $t^{th}$ iteration, the data point $\mathbf{x}_i$ is assigned to the cluster k for the $(t+1)^{th}$ iteration according to:
\begin{equation}
    z_i^{t+1} = \text{argmin}_k \hspace{1mm} ||\mathbf{x}_i - \mu_k^t||_2^2 \notag
\end{equation}
    Where $||\cdot||_2$ is the L2 norm.
    \item Mean Calculation: Once every $\mathbf{x}_i$ has been reassigned in the $t^{th}$ iteration, the respective cluster centers/means $\mu_k^{t+1}$ are to be calculated according to:
\begin{equation}
    \mu_k^{t+1} = \frac{\displaystyle\sum_{i = 1}^{N}\mathbf{1}(z_i^{t+1} = k)\mathbf{x}_i}{\displaystyle\sum_{i = 1}^{N}\mathbf{1}(z_i^{t+1} = k)} \notag
\end{equation}
    Where $\mathbf{1}$ is the proposition function defined as:
\begin{equation}
    \mathbf{1}(z) = 
    \begin{cases}
        1 & \text{if z is true}\\
      0 & \text{otherwise}
    \end{cases}
    \notag
\end{equation}
\end{enumerate}
    The algorithm iterates over steps 2 and 3 until the algorithm converges. Convergence of the Lloyd's algorithm implies that the cluster indicators of every $\mathbf{x}_i$ do not change for further iterations. The Lloyd's algorithm produces a hard clustering for each data point and it always converges to a local minima. Thus, the algorithm is sensitive to the initialisation step and will produce different clusterings based on the initialisation.
\section*{SERIAL LLOYD'S ALGORITHM}
    The serial version of the Lloyd's algorithm can be implemented straightforwardly. The inputs to the program would be the dataset, the number of clusters to be produced, and the total number of observations in the dataset. The convergence criterion can be implemented by calculating an error term according to:
\begin{equation}
    \mathbf{E} = \sum_{i = 1}^{K}||\mu_i^{t+1} - \mu_i^{t}||_2^2 \notag
\end{equation}
    This error term is calculated at the end of each iteration by making use of the cluster centers of two consecutive iterations. This error can be compared with a tolerance value of the order of $10^{-6}$ inside the loop that performs the other steps of the algorithm. Apart from the convergence criterion, the loop will also contain lines of code that will perform Steps 2 and 3 as mentioned above according to the Lloyd's algorithm. Once the algorithm has converged, the final cluster indicators and the corresponding cluster means are produced.

    For evaluating the serial version of the algorithm as well as the parallelised version in the upcoming sections, we shall make use of three datasets. The three datasets are of sizes 100000, 200000, and 500000 and all three of them are generated in a similar manner using a mixture of Bivariate Gaussian Distributions of some mean and covariance. Additionally, we will also be using another set of datasets of three-dimensional points of sizes 100000, 200000, 400000, 800000, and 1 million samples By using datasets of varying sizes, it would allow us to evaluate the algorithm with respect to the scaling of the dataset. 

    The main metric concerning the project is the time complexity of the algorithm as well as the number of iterations required for convergence for various cases. After running the serial algorithm on the datasets for clusters of $k = 4,8,\:\text{and}\:11$, the results are:

\begin{table}[H]
\begin{center}
\begin{tabular}{c  c  c  c}
& & \\ 
\hline
N & K = 4 & K = 8 & K = 11 \\
\hline
500000 (2D) & 1.664616 & 5.313805 & 25.744963 \\

1000000 (3D) & 2.255409 & 34.27957 & 73.925911 \\

\hline
\end{tabular}
\caption{Size of dataset (N) vs time taken for convergence}
\end{center}
\end{table}

\section*{PARALLEL LLOYD'S ALGORITHM}
As seen in the previous section, the Lloyd's algorithm takes an increasing amount of time when the size of the dataset increases, as well as when the number of clusters also increases. However, the main advantage of the Lloyd's algorithm is the underlying simplicity of its steps and calculations. A large portion of the steps in the algorithm can be parallelised, which shall be elucidated in the subsequent subsections of the paper. 
\subsection*{Using OpenMP}

With OpenMP, the aim is to work with a Shared Memory Model on the algorithm. For this particular algorithm, we've majorly implemented a data parallelisation model with task data parallelisation sections. The dataset is to be divided among the number of threads specified by the user. For this, the threads have to be spawned before the algorithm begins. Each thread will independently perform the reassignment step as well as calculate the local cluster means. Once these local cluster means have been calculated, these are transferred to a global variable. The global variable is used by the master thread in calculating the error $\mathbf{E}$ which is then transferred to a global error variable.

For this particular OpenMP program, only the $\mathbf{parallel}$, $\mathbf{critical}$, and $\mathbf{barrier}$ directives have been made use of. The reason why $\mathbf{parallel}$ was preferred over $\mathbf{parallel\;for}$ is because the number of iterations required for convergence is unknown. The $\mathbf{critical}$ directive is used in the global cluster means section to prevent a racing condition among the threads while writing into the global variable.

The results obtained after running the parallelised algorithm for various number of threads for both the two and three-dimensional datasets have been tabulated below. The number of clusters to be produced is fixed to a value of 8 for the 2-dimensional dataset and 4 for the 3-dimensional dataset:

\begin{table}[H]
\begin{center}
\begin{tabular}{c c  c  c  c}
& & \\ 
\hline
N & p = 2 & p = 4 & p = 8 & p = 16 \\
\hline
100000 & 0.680664, & 0.381361, & 0.273247 & 0.378534 \\

200000 & 0.788368 & 0.414747 & 0.324386 & 0.310875 \\

500000 & 10.988341 & 5.538359 & 4.244740 & 3.648641 \\
\hline
\end{tabular}
\caption{2D dataset time taken vs number of threads}
\end{center}
\end{table}

\begin{table}[H]
\begin{center}
\begin{tabular}{c c c c c}
& & \\ 
\hline
N & p = 2 & p = 4 & p = 8 & p = 16 \\
\hline
100000 & 3.33448 & 1.672675 & 1.220420 & 1.230586 \\

200000 & 7.327056 & 3.714937 & 2.728632 & 2.359286 \\

400000 & 14.286552 & 8.2062708 & 6.013132 & 4.937502 \\

800000 & 22.893556 & 13.358098 & 10.931103 & 9.245712 \\

1000000 & 35.973150 & 19.965121 & 16.016981 & 13.495912 \\
\hline
\end{tabular}
\caption{3D dataset time taken vs number of threads}
\end{center}
\end{table}

\subsection*{Using OpenACC}

 With OpenACC, the aim is to create a CPU-GPU work-sharing environment for the algorithm. Since the GPU is intelligent enough to identify the workload, it will appropriately spawn the required number of gangs and workers while running parallel codes. Just like OpenMP, a task parallelisation model is made use of here. The difference between the OpenMP and OpenACC model is that the $\mathbf{parallel}$ directive isn't called at the beginning before the algorithm begins. Rather, the directive is called at the blocks of code corresponding to the various steps of the algorithm. Due to this, there's a constant forking/de-forking of gangs and workers in each iteration, unlike the OpenMP version.

 Since the parallel directives are called for each block of code, this would enable the use of other directives such as $\mathbf{acc\;loop}$,  $\mathbf{parallel\;loop}$, $\mathbf{atomic}$, and $\mathbf{reduction}$ to further parallelise, optimise and speed up the code.

The results obtained after running the parallelised algorithm for both the two and three-dimensional datasets have been tabulated below. The number of clusters to be produced is fixed to a value of 8 for the 2-dimensional dataset and 4 for the 3-dimensional dataset:

\begin{table}[H]
\begin{center}
\begin{tabular}{c  c}
\hline
N & Time Taken \\
\hline
100000 & 0.7213 \\

200000 & 0.283524 \\

500000 &  0.518219 \\
\hline
\end{tabular}
\caption{2D dataset size vs Time Taken}
\end{center}
\end{table}

\begin{table}[H]
\begin{center}
\begin{tabular}{c c}
\hline
N & Time Taken \\
\hline
100000 & 0.087148 \\

200000 & 0.486771 \\

400000 & 0.548548 \\

800000 &  0.743832 \\

1000000 &  0.802407 \\
\hline
\end{tabular}
\caption{3D dataset size vs Time Taken}
\end{center}
\end{table}
\section*{RESULTS AND DISCUSSION}
The results of clustering the 3-dimensional dataset into 4 clusters have been plotted below for both the serial and the parallel program by OpenACC. We can observe that the parallel program achieves similar clustering to the serial program which also are the optimal clusters for K = 4.
\begin{figure}[hbt!]
    \centering
    \includegraphics[width = 0.3\textwidth]{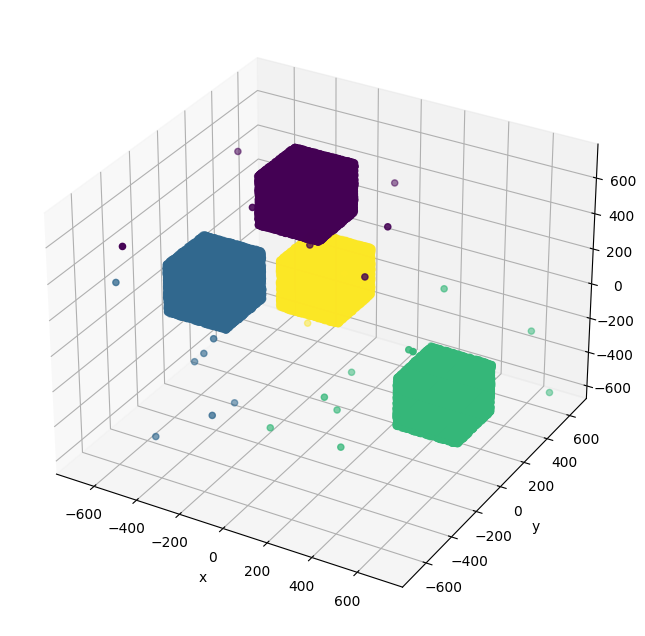}
    \caption{Results of Serial K-Means on 1M datapoints}
    \label{fig:1M-serial}
\end{figure}
\begin{figure}[hbt!]
    \centering
    \includegraphics[width = 0.3\textwidth]{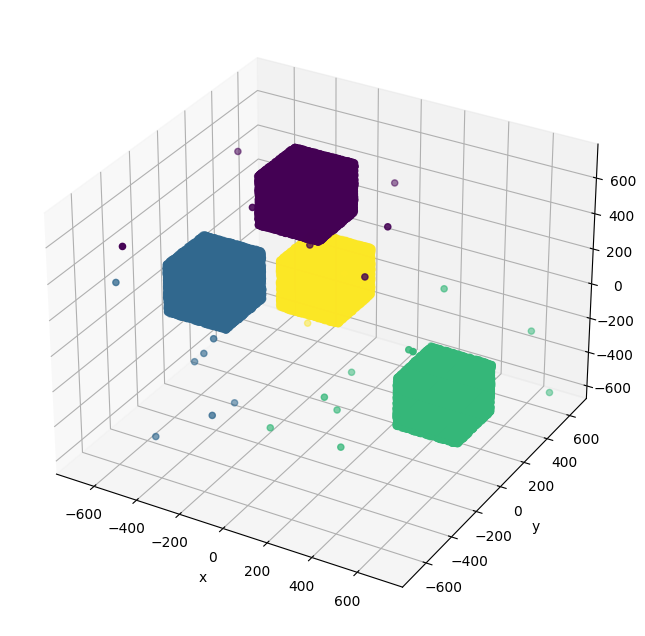}
    \caption{Results of Parallel K-Means on 1M datapoints}
    \label{fig:1M-parallel}
\end{figure}
\begin{figure}[hbt!]
    \centering
    \includegraphics[width = 0.3\textwidth]{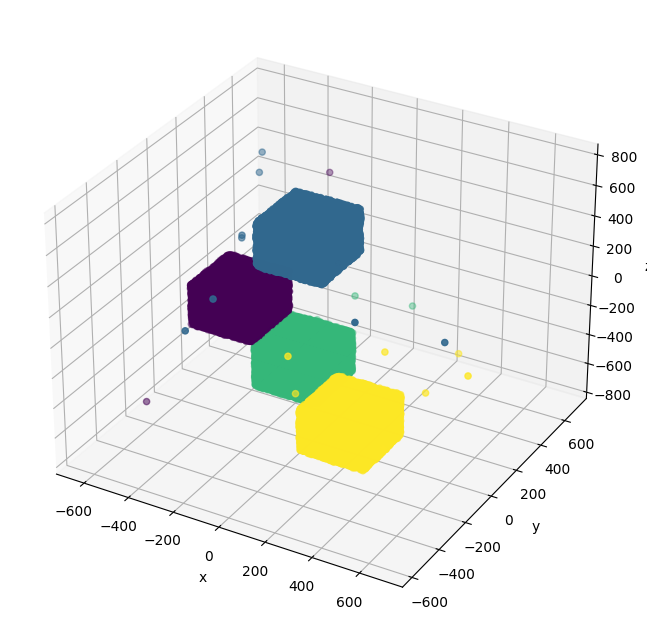}
    \caption{Results of Serial K-Means on 400k datapoints}
    \label{fig:400k}
\end{figure}
\begin{figure}[hbt!]
    \centering
    \includegraphics[width = 0.3\textwidth]{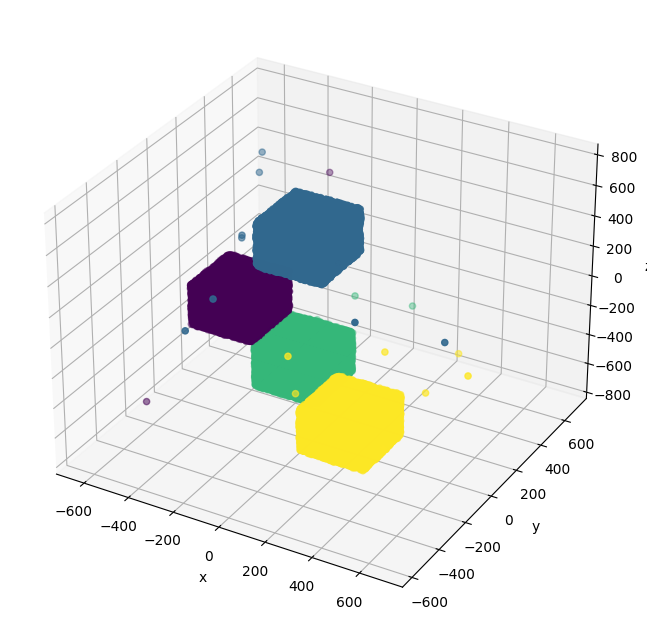}
    \caption{Results of parallel K-Means on 400k datapoints}
    \label{fig:400k-parallel}
\end{figure}
\par
The results of clustering the 2-dimensional dataset into 11 clusters have been plotted below for both the serial and the parallel program by OpenACC. The parallel program achieves similar clustering as the serial program. The clusters may not be optimal due to overlapping regions between them, the presence of closely spaced groups of points and possibly needing more iterations to generate more optimal clusters due to a large number of clusters.
\begin{figure}[hbt!]
    \centering
    \includegraphics[width = 0.3\textwidth]{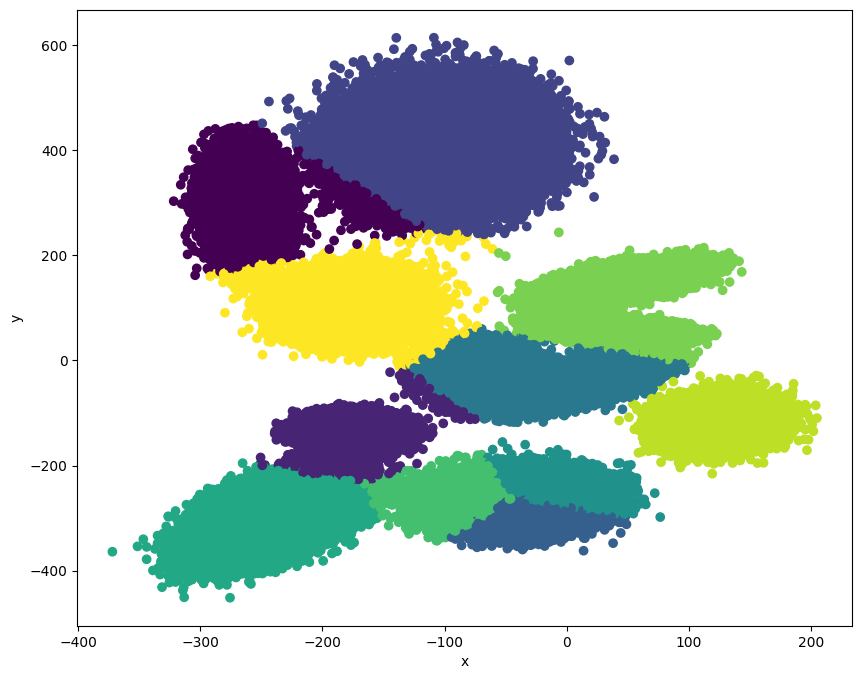}
    \caption{Results of Serial K-Means on 500k 2D Dataset}
    \label{fig:500k}
\end{figure}
\begin{figure}[hbt!]
    \centering
    \includegraphics[width = 0.3\textwidth]{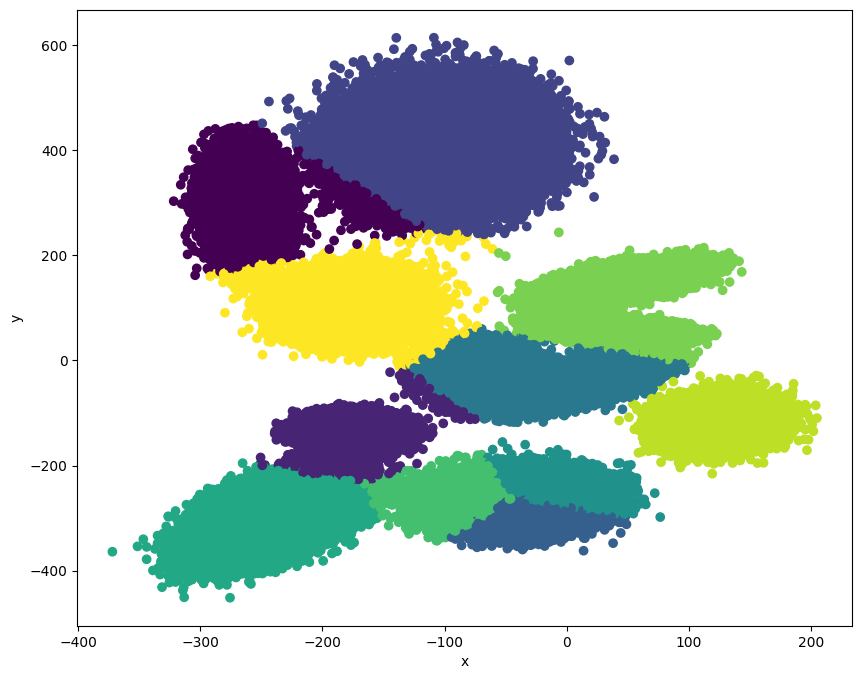}
    \caption{Results of Parallel K-Means on 500k 2D Dataset}
    \label{fig:500k-2d}
\end{figure}
\par
The Speedup $\psi(n,p)$ as a function of number of threads ($p$) for both the 2D and 3D datasets has been plotted as well. We can see an increase in the speedup as the size of the dataset increases across the 2D and 3D datasets (except for maybe the smallest datasets due to the small size of the datasets). We can also observe that the speedup values are larger for the larger datasets, indicating strongly that for large datasets parallelization can offer a significant boost.
\begin{figure}[hbt!]
    \centering
    \includegraphics[width = 0.35\textwidth]{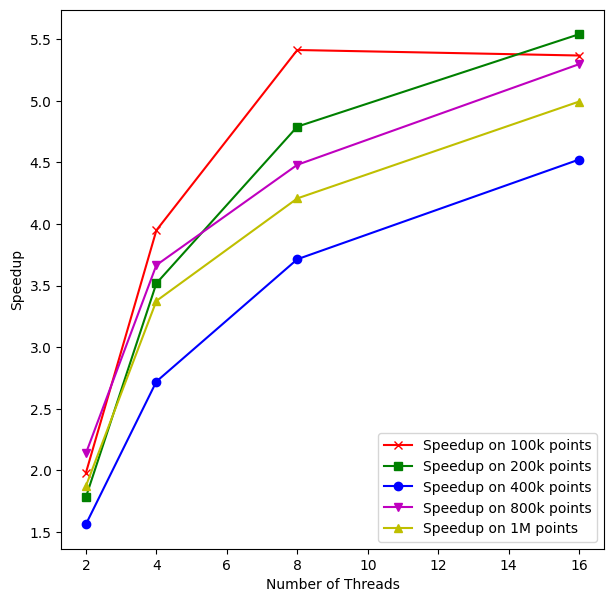}
    \caption{Speedup for 3D Dataset}
    \label{fig:speedup3d}
\end{figure}
\begin{figure}[hbt!]
    \centering
    \includegraphics[width = 0.35\textwidth]{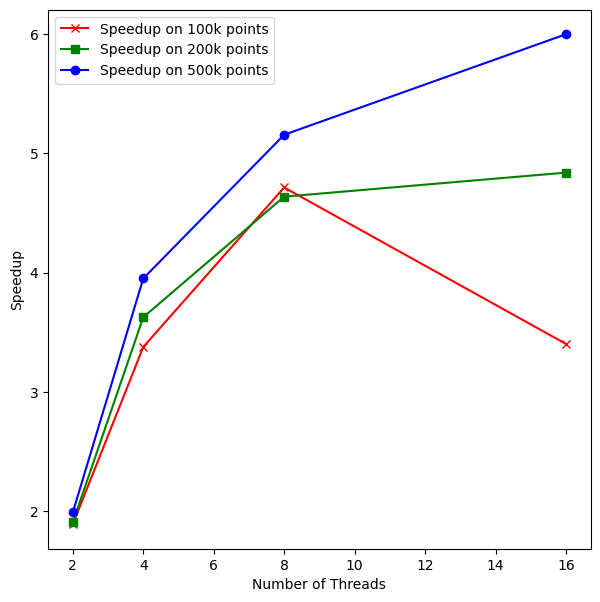}
    \caption{Speedup for 2D Dataset}
    \label{fig:speedup2D}
\end{figure}
\par
The Efficiency $\epsilon(n,p)$ as a function of number of threads ($p$) for both the 2D and 3D datasets has been plotted too. We can observe that the highest efficiency occurs for the number of threads = 2 and it drops as the number of threads increases.
\begin{figure}[hbt!]
    \centering
    \includegraphics[width = 0.4\textwidth]{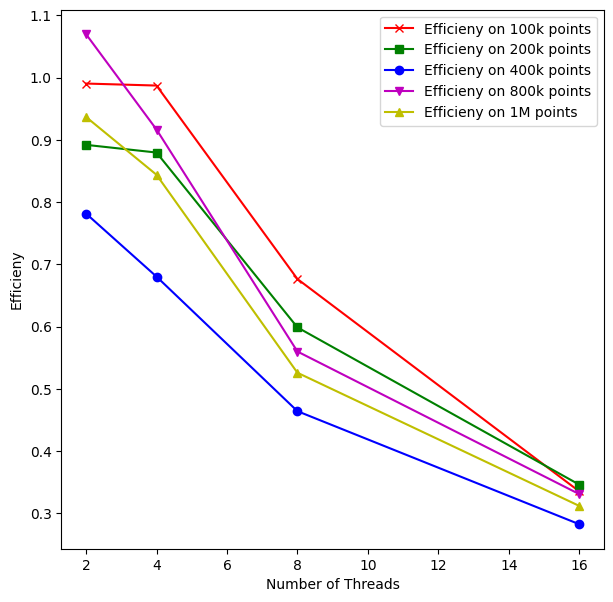}
    \caption{Efficiency for 3D Dataset}
    \label{fig:efficiency3d}
\end{figure}
\begin{figure}[hbt!]
    \centering
    \includegraphics[width=0.4\textwidth]{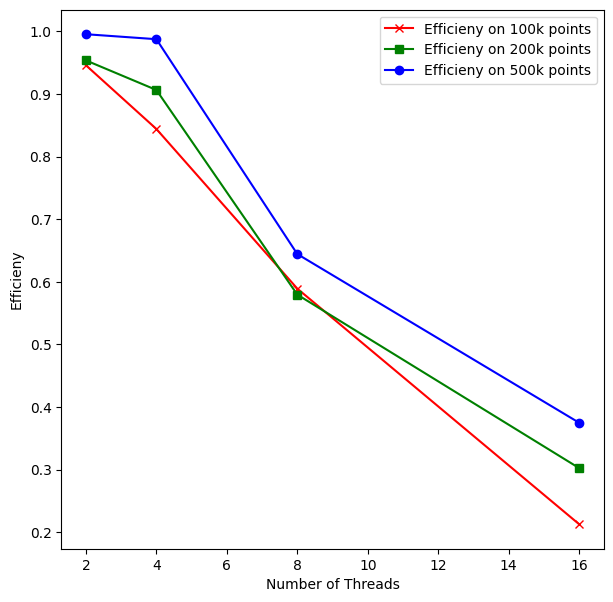}
    \caption{Efficiency for 2D Dataset}
    \label{fig:efficiency2d}
\end{figure}
\par
The variation of time taken with the scaling/size of the dataset for both the 2D and 3D datasets. We can see that for the same value of K, the time taken to compute the clusters increases as the size of the dataset increases.
\begin{figure}[hbt!]
    \centering
    \includegraphics[width = 0.4\textwidth]{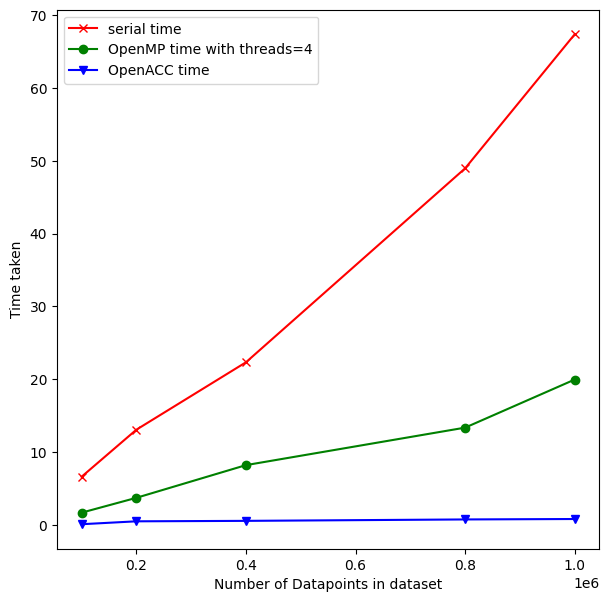}
    \caption{Time taken vs Scaling for 3D Datasets}
    \label{fig:time}
\end{figure}
\begin{figure}[hbt!]
    \centering
    \includegraphics[width=0.4\textwidth]{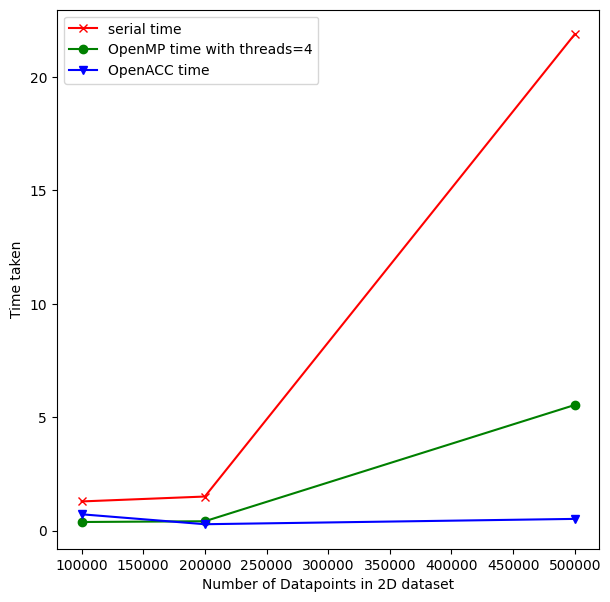}
    \caption{Time taken vs Scaling for 2D Datasets}
    \label{fig:time2d}
\end{figure}

\section*{CONCLUSIONS}
With this paper, we have successfully implemented a parallelised version of the K-Means algorithm with both a shared memory model as well as with GPU programming. Both parallel versions were able to produce results with no loss in accuracy and an appreciable decrease in computation time. When comparing both the OpenMP and OpenACC models, it is observed that the OpenACC version performs better in terms of saving computation time. This provides scope for decreasing computation time in extremely large datasets with real-world data or complex applications of the clustering paradigm such as image segmentation, anomaly detection, etc.
\cite{paper-1}
\cite{paper-2}
\cite{paper-3}
\cite{paper-4}
\bibliographystyle{plain}
\bibliography{asme2e.bib}




%
% ...




\end{document}